%
%
\def\unredoffs{} \def\redoffs{\voffset=-.31truein\hoffset=-.59truein}
\def\speclscape{}
%
%
%
%
\newbox\leftpage \newdimen\fullhsize \newdimen\hstitle \newdimen\hsbody
\tolerance=1000\hfuzz=2pt
\catcode`\@=11 
\def\bigans{b }
\def\answ{b }
%

\ifx\answ\bigans\message{(This will come out unreduced.}
\magnification=1200\unredoffs\baselineskip=16pt plus 2pt minus 1pt
\hsbody=\hsize \hstitle=\hsize 
\else\message{(This will be reduced.} \let\l@r=L
\magnification=1000\baselineskip=16pt plus 2pt minus 1pt \vsize=7truein
\redoffs \hstitle=8truein\hsbody=4.75truein\fullhsize=10truein\hsize=\hsbody
\output={\ifnum\pageno=0 
  \shipout\vbox{\speclscape{\hsize\fullhsize\makeheadline}
   \hbox to \fullhsize{\hfill\pagebody\hfill}}\advancepageno
  \else
 \almostshipout{\leftline{\vbox{\pagebody\makefootline}}}\advancepageno 
  \fi}
\def\almostshipout#1{\if L\l@r \count1=1 \message{[\the\count0.\the\count1]}
      \global\setbox\leftpage=#1 \global\let\l@r=R
 \else \count1=2
  \shipout\vbox{\speclscape{\hsize\fullhsize\makeheadline}
      \hbox to\fullhsize{\box\leftpage\hfil#1}}  \global\let\l@r=L\fi}
\fi
%
\newcount\yearltd\yearltd=\year\advance\yearltd by -1900

\def\Title#1#2{\nopagenumbers\abstractfont\hsize=\hstitle\rightline{#1}%
\vskip 1in\centerline{\titlefont #2}\abstractfont\vskip .5in\pageno=0}
\def\Date#1{\vfill\leftline{#1}\tenpoint\supereject\global\hsize=\hsbody%
\footline={\hss\tenrm\folio\hss}}
%

\def\draftmode{\message{ DRAFTMODE }\def\draftdate{{\rm preliminary draft:
\number\month/\number\day/\number\yearltd\ \ \hourmin}}%
\headline={\hfil\draftdate}\writelabels\baselineskip=20pt plus 2pt minus 2pt
 {\count255=\time\divide\count255 by 60 \xdef\hourmin{\number\count255}
  \multiply\count255 by-60\advance\count255 by\time
  \xdef\hourmin{\hourmin:\ifnum\count255<10 0\fi\the\count255}}}
\def\nolabels{\def\wrlabeL##1{}\def\eqlabeL##1{}\def\reflabeL##1{}}
\def\writelabels{\def\wrlabeL##1{\leavevmode\vadjust{\rlap{\smash%
{\line{{\escapechar=` \hfill\rlap{\sevenrm\hskip.03in\string##1}}}}}}}%
\def\eqlabeL##1{{\escapechar-1\rlap{\sevenrm\hskip.05in\string##1}}}%
\def\reflabeL##1{\noexpand\llap{\noexpand\sevenrm\string\string\string##1}}}
\nolabels
%
\global\newcount\secno \global\secno=0
\global\newcount\meqno \global\meqno=1
\def\newsec#1{\global\advance\secno by1\message{(\the\secno. #1)}
\global\subsecno=0\eqnres@t\noindent{\bf\the\secno. #1}
\writetoca{{\secsym} {#1}}\par\nobreak\medskip\nobreak}
\def\eqnres@t{\xdef\secsym{\the\secno.}\global\meqno=1\bigbreak\bigskip}
\def\sequentialequations{\def\eqnres@t{\bigbreak}}\xdef\secsym{}
\global\newcount\subsecno \global\subsecno=0
\def\subsec#1{\global\advance\subsecno by1\message{(\secsym\the\subsecno. #1)}
\ifnum\lastpenalty>9000\else\bigbreak\fi
\noindent{\it\secsym\the\subsecno. #1}\writetoca{\string\quad 
{\secsym\the\subsecno.} {#1}}\par\nobreak\medskip\nobreak}
\def\appendix#1#2{\global\meqno=1\global\subsecno=0\xdef\secsym{\hbox{#1.}}
\bigbreak\bigskip\noindent{\bf Appendix #1. #2}\message{(#1. #2)}
\writetoca{Appendix {#1.} {#2}}\par\nobreak\medskip\nobreak}
%
%
\def\eqnn#1{\xdef #1{(\secsym\the\meqno)}\writedef{#1\leftbracket#1}%
\global\advance\meqno by1\wrlabeL#1}
\def\eqna#1{\xdef #1##1{\hbox{$(\secsym\the\meqno##1)$}}
\writedef{#1\numbersign1\leftbracket#1{\numbersign1}}%
\global\advance\meqno by1\wrlabeL{#1$\{\}$}}
\def\eqn#1#2{\xdef #1{(\secsym\the\meqno)}\writedef{#1\leftbracket#1}%
\global\advance\meqno by1$$#2\eqno#1\eqlabeL#1$$}
%
\newskip\footskip\footskip14pt plus 1pt minus 1pt 
\def\footnotefont{\ninepoint}\def\f@t#1{\footnotefont #1\@foot}
\def\f@@t{\baselineskip\footskip\bgroup\footnotefont\aftergroup\@foot\let\next}
\setbox\strutbox=\hbox{\vrule height9.5pt depth4.5pt width0pt}
\global\newcount\ftno \global\ftno=0
\def\foot{\global\advance\ftno by1\footnote{$^{\the\ftno}$}}
%
\newwrite\ftfile   
\def\footend{\def\foot{\global\advance\ftno by1\chardef\wfile=\ftfile
$^{\the\ftno}$\ifnum\ftno=1\immediate\openout\ftfile=foots.tmp\fi%
\immediate\write\ftfile{\noexpand\smallskip%
\noexpand\item{f\the\ftno:\ }\pctsign}\findarg}%
\def\footatend{\vfill\eject\immediate\closeout\ftfile{\parindent=20pt
\centerline{\bf Footnotes}\nobreak\bigskip\input foots.tmp }}}
\def\footatend{}
%
%
\global\newcount\refno \global\refno=1
\newwrite\rfile
\def\ref{[\the\refno]\nref}
\def\nref#1{\xdef#1{[\the\refno]}\writedef{#1\leftbracket#1}%
\ifnum\refno=1\immediate\openout\rfile=refs.tmp\fi
\global\advance\refno by1\chardef\wfile=\rfile\immediate
\write\rfile{\noexpand\item{#1\ }\reflabeL{#1\hskip.31in}\pctsign}\findarg}
\def\findarg#1#{\begingroup\obeylines\newlinechar=`\^^M\pass@rg}
{\obeylines\gdef\pass@rg#1{\writ@line\relax #1^^M\hbox{}^^M}%
\gdef\writ@line#1^^M{\expandafter\toks0\expandafter{\striprel@x #1}%
\edef\next{\the\toks0}\ifx\next\em@rk\let\next=\endgroup\else\ifx\next\empty%
\else\immediate\write\wfile{\the\toks0}\fi\let\next=\writ@line\fi\next\relax}}
\def\striprel@x#1{} \def\em@rk{\hbox{}} 
\def\lref{\begingroup\obeylines\lr@f}
\def\lr@f#1#2{\gdef#1{\ref#1{#2}}\endgroup\unskip}

\def\addref#1{\immediate\write\rfile{\noexpand\item{}#1}} 
\def\startrefs#1{\immediate\openout\rfile=refs.tmp\refno=#1}
\def\xref{\expandafter\xr@f}\def\xr@f[#1]{#1}
\def\refs#1{\count255=1[\r@fs #1{\hbox{}}]}
\def\r@fs#1{\ifx\und@fined#1\message{reflabel \string#1 is undefined.}%
\nref#1{need to supply reference \string#1.}\fi%
\vphantom{\hphantom{#1}}\edef\next{#1}\ifx\next\em@rk\def\next{}%
\else\ifx\next#1\ifodd\count255\relax\xref#1\count255=0\fi%
\else#1\count255=1\fi\let\next=\r@fs\fi\next}
%

%
\newwrite\ffile\global\newcount\figno \global\figno=1
\def\fig{Figure~\the\figno\nfig}
\def\nfig#1{\xdef#1{Figure~\the\figno}%
\writedef{#1\leftbracket fig.\noexpand~\the\figno}%
\ifnum\figno=1\immediate\openout\ffile=figs.tmp\fi\chardef\wfile=\ffile%
\immediate\write\ffile{\noexpand\medskip\noexpand\item{Fig.\ \the\figno. }
\reflabeL{#1\hskip.55in}\pctsign}\global\advance\figno by1\findarg}
\def\xfig{\expandafter\xf@g}\def\xf@g fig.\penalty\@M\ {}
\def\figs#1{figs.~\f@gs #1{\hbox{}}}
\def\f@gs#1{\edef\next{#1}\ifx\next\em@rk\def\next{}\else
\ifx\next#1\xfig #1\else#1\fi\let\next=\f@gs\fi\next}
\newwrite\lfile
{\escapechar-1\xdef\pctsign{\string\%}\xdef\leftbracket{\string\{}
\xdef\rightbracket{\string\}}\xdef\numbersign{\string\#}}

\def\writestop{\def\writestoppt{\immediate\write\lfile{\string\pageno%
\the\pageno\string\startrefs\leftbracket\the\refno\rightbracket%
\string\def\string\secsym\leftbracket\secsym\rightbracket%
\string\secno\the\secno\string\meqno\the\meqno}\immediate\closeout\lfile}}
\def\writestoppt{}\def\writedef#1{}
\def\seclab#1{\xdef #1{\the\secno}\writedef{#1\leftbracket#1}\wrlabeL{#1=#1}}
\def\subseclab#1{\xdef #1{\secsym\the\subsecno}%
\writedef{#1\leftbracket#1}\wrlabeL{#1=#1}}
\newwrite\tfile \def\writetoca#1{}
\def\leaderfill{\leaders\hbox to 1em{\hss.\hss}\hfill}
\def\writetoc{\immediate\openout\tfile=toc.tmp 
   \def\writetoca##1{{\edef\next{\write\tfile{\noindent ##1 
   \string\leaderfill {\noexpand\number\pageno} \par}}\next}}}

%
\catcode`\@=12 
%
\edef\tfontsize{\ifx\answ\bigans scaled\magstep3\else scaled\magstep4\fi}
\font\titlerm=cmr10 \tfontsize \font\titlerms=cmr7 \tfontsize
\font\titlermss=cmr5 \tfontsize \font\titlei=cmmi10 \tfontsize
\font\titleis=cmmi7 \tfontsize \font\titleiss=cmmi5 \tfontsize
\font\titlesy=cmsy10 \tfontsize \font\titlesys=cmsy7 \tfontsize
\font\titlesyss=cmsy5 \tfontsize \font\titleit=cmti10 \tfontsize
\skewchar\titlei='177 \skewchar\titleis='177 \skewchar\titleiss='177
\skewchar\titlesy='60 \skewchar\titlesys='60 \skewchar\titlesyss='60
\def\titlefont{\def\rm{\fam0\titlerm}
\textfont0=\titlerm \scriptfont0=\titlerms \scriptscriptfont0=\titlermss
\textfont1=\titlei \scriptfont1=\titleis \scriptscriptfont1=\titleiss
\textfont2=\titlesy \scriptfont2=\titlesys \scriptscriptfont2=\titlesyss
\textfont\itfam=\titleit \def\it{\fam\itfam\titleit}\rm}
 \ifx\answ\bigans\else scaled\magstep1\fi
\ifx\answ\bigans\def\abstractfont{\tenpoint}\else
\font\abssl=cmsl10 scaled \magstep1
\font\absrm=cmr10 scaled\magstep1 \font\absrms=cmr7 scaled\magstep1
\font\absrmss=cmr5 scaled\magstep1 \font\absi=cmmi10 scaled\magstep1
\font\absis=cmmi7 scaled\magstep1 \font\absiss=cmmi5 scaled\magstep1
\font\abssy=cmsy10 scaled\magstep1 \font\abssys=cmsy7 scaled\magstep1
\font\abssyss=cmsy5 scaled\magstep1 \font\absbf=cmbx10 scaled\magstep1
\skewchar\absi='177 \skewchar\absis='177 \skewchar\absiss='177
\skewchar\abssy='60 \skewchar\abssys='60 \skewchar\abssyss='60
\def\abstractfont{\def\rm{\fam0\absrm}
\textfont0=\absrm \scriptfont0=\absrms \scriptscriptfont0=\absrmss
\textfont1=\absi \scriptfont1=\absis \scriptscriptfont1=\absiss
\textfont2=\abssy \scriptfont2=\abssys \scriptscriptfont2=\abssyss
\textfont\itfam=\bigit \def\it{\fam\itfam\bigit}\def\footnotefont{\tenpoint}%
\textfont\slfam=\abssl \def\sl{\fam\slfam\abssl}%
\textfont\bffam=\absbf \def\bf{\fam\bffam\absbf}\rm}\fi
\def\tenpoint{\def\rm{\fam0\tenrm}
\textfont0=\tenrm \scriptfont0=\sevenrm \scriptscriptfont0=\fiverm
\textfont1=\teni  \scriptfont1=\seveni  \scriptscriptfont1=\fivei
\textfont2=\tensy \scriptfont2=\sevensy \scriptscriptfont2=\fivesy
\textfont\itfam=\tenit \def\it{\fam\itfam\tenit}\def\footnotefont{\ninepoint}%
\textfont\bffam=\tenbf \def\bf{\fam\bffam\tenbf}\def\sl{\fam\slfam\tensl}\rm}
\font\ninerm=cmr9 \font\sixrm=cmr6 \font\ninei=cmmi9 \font\sixi=cmmi6 
\font\ninesy=cmsy9 \font\sixsy=cmsy6 \font\ninebf=cmbx9 
\font\nineit=cmti9 \font\ninesl=cmsl9 \skewchar\ninei='177
\skewchar\sixi='177 \skewchar\ninesy='60 \skewchar\sixsy='60 
\def\ninepoint{\def\rm{\fam0\ninerm}
\textfont0=\ninerm \scriptfont0=\sixrm \scriptscriptfont0=\fiverm
\textfont1=\ninei \scriptfont1=\sixi \scriptscriptfont1=\fivei
\textfont2=\ninesy \scriptfont2=\sixsy \scriptscriptfont2=\fivesy
\textfont\itfam=\ninei \def\it{\fam\itfam\nineit}\def\sl{\fam\slfam\ninesl}%
\textfont\bffam=\ninebf \def\bf{\fam\bffam\ninebf}\rm} 
%
%

\hyphenation{anom-aly anom-alies coun-ter-term coun-ter-terms}
\def\inv{^{\raise.15ex\hbox{${\scriptscriptstyle -}$}\kern-.05em 1}}

\def\Dsl{\,\raise.15ex\hbox{/}\mkern-13.5mu D} 
\def\dsl{\raise.15ex\hbox{/}\kern-.57em\partial}

\font\bigit=cmti10 scaled \magstep1
\def\lspace{\ifx\answ\bigans{}\else\qquad\fi}
\def\lbspace{\ifx\answ\bigans{}\else\hskip-.2in\fi} 
\def\boxeqn#1{\vcenter{\vbox{\hrule\hbox{\vrule\kern3pt\vbox{\kern3pt
	\hbox{${\displaystyle #1}$}\kern3pt}\kern3pt\vrule}\hrule}}}
\def\mbox#1#2{\vcenter{\hrule \hbox{\vrule height#2in
		\kern#1in \vrule} \hrule}}  
%
    
 \def\CH{{\cal H}}

\def\grad#1{\,\nabla\!_{{#1}}\,}

\def\darr#1{\raise1.5ex\hbox{$\leftrightarrow$}\mkern-16.5mu #1}

\def\half{{\textstyle{1\over2}}} 
\def\roughly#1{\raise.3ex\hbox{$#1$\kern-.75em\lower1ex\hbox{$\sim$}}}

\def\nv{{\bf n}}
\def\Nv{{\bf N}}

\def\gradv{\grad}

\def\half{{1\over 2}}

\def\kbT{k_{\scriptscriptstyle\rm B}T}

\def\bold#1{\setbox0=\hbox{$#1$}%
     \kern-.010em\copy0\kern-\wd0
     \kern.025em\copy0\kern-\wd0
     \kern-.020em\raise.0200em\box0 }
\def\dnb{\delta{\bf n}}

\def\grad{\bold{\nabla}}

\lref\NLS{
T.~Chan, C.W.~Garland and H.T.~Nguyen, Phys. Rev. E, {\bf 52} (1995) 5000; L.
Navailles, unpublished (1995).}

\lref\NS{
D.R.~Nelson, Phys. Rev. Lett. {\bf60} (1988) 1973;
D.R.~Nelson and H.S. Seung, Phys. Rev. B {\bf 39} (1989) 9153.}

\lref\SEL{J.V.~Selinger and J.M.~Schnur, Phys. Rev. Lett. {\bf 71}
(1993) 4091; J.V. Selinger, Z.-G.~Wang, R.F.~Bruinsma and C.M.~Knobler,
Phys. Rev. Lett. {\bf 70} (1993) 1139.}

\lref\PN{P.~Nelson and T.~Powers, Phys. Rev. Lett. {\bf 69} (1992) 3409;
J. Phys. II (Paris) {\bf 3} (1993) 1535.}

\lref\SL{S.~Langer and J.~Sethna, Phys. Rev. A {\bf 34} (1986) 5035;
G.A.~Hinshaw, R.G.~Petschek and R.A.~Pelcovits,
Phys. Rev. Lett. {\bf 60} (1988) 1864.}

\lref\dg{P.G.~de~Gennes, Solid State Commun. {\bf 14} (1973) 997.}

\lref\helfrich{W.~Helfrich, Z. Naturforsch. {\bf 28C} (1973) 693;
P.~Canham, J. Theor. Biol. {\bf 26} (1970) 61.}

\lref\NelsonPowers{P.~Nelson and T.~Powers, Phys. Rev. Lett. {\bf 69}
(1992) 3409; J. Phys. II (Paris) {\bf 3} (1993) 1535.}

\lref\LubKam{R.D.~Kamien and T.C.~Lubensky, J. Phys. I (Paris) {\bf 3}
(1993) 2131.}

\lref\lks{T.C.~Lubensky, R.D.~Kamien and H.~Stark, to appear in
Mol. Cryst. Liq. Cryst. (1996) [cond-mat/9512163].}

\lref\NT{W.~Helfrich, J. Phys. (Paris) {\bf 39} (1978) 1199;
D.R.~Nelson and J.~Toner, Phys. Rev. B {\bf 24} (1981) 363.}

\Title{}{\vbox{\centerline{Chiral Lyotropic Liquid
Crystals:}\vskip2pt\centerline{
TGB Phases and Helicoidal Structures}}}

\centerline{Randall D. Kamien and T.C. Lubensky}
\smallskip\centerline{\sl Department of Physics and Astronomy, University
of Pennsylvania, Philadelphia, PA 19104}

\vskip .3in
The molecules in lyotropic membranes are typically aligned with the surface
normal.  When these molecules are chiral, there is a tendency for the
molecular direction to twist.  These competing effects can reach a
compromise by producing helicoidal defects in the membranes.  Unlike
thermotropic smectics, the centers of these defects are hollow and
thus their energy cost comes from the line energy of an exposed lamellar
surface.
We describe both the twist-grain-boundary phase of chiral lamellar phases
as well as the isolated helicoidal defects.

\Date{20 May 1996}

\newsec{Introduction}
Lyotropic liquid crystals share a number of features with their thermotropic
cousins.
In lyotropic lamellar systems, however, it is believed that because the
molecules are amphiphillic the molecular axes will align with the layer normal
in
an untilted, $L_\alpha$ phase.  Chiral molecules are well known to exert
torques on each
other \ref\HKL{A.B.~Harris, R.D.~Kamien and T.C.~Lubensky, {\sl in preparation}
(1996).}
leading to chiral mesophases in thermotropic liquid crystals.  In chiral
lyotropics then,
there will be a frustration between normal alignment and the tendency of the
molecules to twist \lks.

In thermotropic smectic-$A$ phases, this frustration can be relieved via the
formation of a
twist-grain-boundary phase (TGB) \ref\LR{S.R.~Renn and T.C.~Lubensky, Phys.
Rev. A {\bf 38} (1988) 2132; {\bf 41} (1990) 4392.}\
analogous to the Abrikosov flux line lattice of type-II
superconductors.
In this case the defects are screw dislocations with cores made of the
associated
nematic liquid
crystal phase.  In this letter we will describe a similar screw dislocation
structure
for lyotropic lamellae and propose an $L_\alpha$ TGB phase.  We will also
discuss
a possible defect mediated phase transition \NT\ between the $L_\alpha$ phase
and
a normal cholesteric phase \ref\ROUX{D.~Roux, {\sl private communication}
(1995).}.

\newsec{Helicoidal Defects in Membranes}
We model the free energy of an isolated bilayer membrane as a sum of
contributions.
The free energy for membrane fluctuations is \ref\CH{
W.~Helfrich, Z. Naturforsch. {\bf 28C} (1973) 693;
P.~Canham, J. Theor. Biol. {\bf 26} (1970) 61.}
\eqn\efreeh{F_m = {\kappa\over 2}\int dS\,\left({1 \over R_1} + {1 \over
R_2}\right)^2 = {\kappa\over 2} \int dS\, ( \gradv \cdot \Nv)^2,
}
where $R_1$ and $R_2$ are the principal curvatures, $\Nv$ is the
layer normal and $\gradv\cdot\Nv$ is the mean curvature.
The free energy for fluctuations of the molecular axis $\nv$ is
given by the membrane version of the chiral Frank free energy, namely:
\eqn\efreen{\eqalign{
F^*_n = {1\over 2}\int dS\,&\bigg\{K_1 (\gradv_{\perp} \cdot \nv )^2 + K_2
(\nv\cdot\gradv_{\perp} \times \nv - q_0 )^2\cr&\qquad
+K_3\left[\nv\times\left(\gradv_{\perp}\times
\nv\right)\right]^2\bigg\},\cr
}}
where $\gradv_{\perp}$ refers to a gradient in the tangent plane of the
membrane.  At quadratic
order, the non-linear terms arising from properly constructed, covariant
derivatives do
not come in.
We couple these two free energies together by adding a term which favors the
alignment of the layer normal with the molecular axis:
\eqn\efreenm{F_{mn}=\gamma\int dS\,
\left\{1-\left(\nv\cdot\Nv\right)^2\right\}.}
The total free energy is the sum $F=F_m+F^*_n + F_{mn}$.
In the Monge gauge, we may represent the surface as a height function $h(x,y)$
over
the co\"ordinates $x$ and $y$.  In this case the layer normal is
$\Nv=\left[-\partial_x h,-\partial_y h,1\right]/\sqrt{1+(\nabla_\perp h)^2}$.
Expanding
the nematic director about the ground state direction $\nv = \left({\bf\hat z}
+
\dnb
\right)/\sqrt{1+\delta n^2}$,
the free energy to quadratic order in the deviations from
the ground state becomes
\eqn\efreemii{F = \half\int dxdy\,\left\{\gamma
\left(\nabla_\perp h + \dnb\right)^2 + \kappa\left(\nabla_\perp^2
h\right)^2\right\}
+F^*_n.}
Note that if we require $\dnb = -\gradv_{\perp}h$ (which follows when
$\gamma\rightarrow\infty$),
or, in other words, if we require $\nv||\Nv$, the resulting effective free
energy has
no chiral term since $\gradv_\perp\times\dnb=0$.  It is only by relaxing the
constraint that
$\nv$ be parallel to $\Nv$ that we see manifestations of chirality.  Thus, free
energies of the $L_\alpha$
phase of membranes (with $\nv||\Nv$) based only on the local membrane geometry
will never
show the effects of chiral molecules.  The $L_\alpha$ phase is the analog of
the smectic-$A$
phase in which all twist and bend have been expelled \dg.

In the $L_{\beta'}$ phases, there is an equilibrium projection $\bf c$ of the
the director $\nv$ onto
the tangent plane, and there are manifestations of chirality in the free energy
involving
${\bf c}$ \refs{\SL,\SEL,\PN}, for example, through a term of the form $\int
{\bf c}\cdot\gradv_{\perp}\times{\bf c}$.
These terms are responsible for a number of interesting modulated structures in
and of the membrane.

The free energy \efreemii\ is very similar to that for smectic-$A$ liquid
crystals,
although the functions are restricted to depend only on the membrane
co\"ordinates.
If we consider a screw-like configuration of the surface $h(x,y)=b\arctan(y/x)$
(where
$b$ is the ``Burgers vector'' of the defect), then
we can calculate the energy of the associated configuration, after relaxing
${\bf\delta n}
0=\nabla_\perp h$ ({\sl i.e.}, $\nv=\Nv$).  This configuration, however, is
singular
at the origin and is not allowed.  To facilitate the screw-like defect the
membrane
must cease to exist in the core, or, in other words, there will be exposed
membrane
edges
around a solvent-filled cylindrical core. Thus, in the presence of a screw
defect
we must now add a line energy.  This line energy comes from two pieces: the
first
is the energy cost of having an exposed edge, while the second comes from any
surface
terms in the free energy, for instance the Gaussian curvature.  We group these
energies
together into a term:
\eqn\efreeedge{F_e = \mu\int_C d\ell,}
where $C$ is the inner edge of the helicoidal surface.

Far away from the cylindrical core,
$\dnb=b\left[y,-x,0\right]/(x^2+y^2)$.  Between
the core radius $\xi$ and the twist penetration depth $\lambda =
\sqrt{K_2/\gamma}$,
the
nematic director relaxes to its large radius value.  This
means that the core has a double-twist texture, as found in liquid crystalline
blue phases.
A perfect helicoidal surface has no mean curvature, thus our defect does
not contribute to $F_m$.  The free energies per turn
are (with $\lambda=\sqrt{K_2/\gamma}$):
\eqn\efrees{\eqalign{
F_{mn} & = \half\gamma b^2\pi\ln\left[1+{\lambda^2\over\xi^2}\right]\cr
F_n &  =-K_2q_0b \cr
F_e & = \mu\sqrt{b^2+(2\pi\xi)^2},\cr}}
where, in the London (type II) limit, $\xi\ll\lambda$.  To minimize the sum,
note that
only $F_n$ can be negative: this chooses the sign of $b$.
Minimizing the sum of the constituent free energies, we find that there
will always be a non-zero value of $b$ for which the energy is minimized.
In the two limits ($b\ll\xi$ or $b\gg\xi$) we find:
\eqn\elimits{
b^*=\left\{\matrix{
{\displaystyle 2\pi\xi K_2q_0\over
\displaystyle 2\pi^2\xi\gamma\ln\left(\lambda^2/\xi^2+1\right)+\mu}
&\quad b^*\ll\xi\cr
{}~&~\cr~&~\cr
{\displaystyle K_2q_0-\mu\over\displaystyle
\pi\gamma\ln\left(\lambda^2/\xi^2+1\right)}
\hfill &\quad
b^*\gg\xi .\cr}\right.}
At the same time, $\xi$ is determined also via minimizing the free energy.
We thus have
\eqn\exi{\gamma^2\lambda^4(b^*)^4\left[(b^*)^2+(2\pi\xi^*)^2\right]
= 16\pi^2\mu^2\left[\lambda^2+
(\xi^*)^2\right]^2(\xi^*)^4.}
The optimal value of $\xi=\xi^*$ will determine the size of the hole in the
membrane.

Unlike a layered system where $b$ is
quantized
in units of the spacing, $b$ varies continuously
with $K_2q_0$.  When a defect
appears, the membrane must continue wrapping around into a many-layered
helicoidal surface.  Not to do so would create a line of exposed molecules
leading out from the center of the defect $r=0$ to the boundary $r=\infty$.

The line energy $\mu$ presumably contains both energetic and entropic factors.
In principle, $\mu$ can be negative at the temperatures we are considering.  If
that were the case, even the non-chiral membrane, made of racemic mixture of
molecules would unbind, leading to a randomly defective, layered structure.  A
lyotropic with a negative line tension would never really form a single layer
structure and would have defects proliferating at the molecular level leading
to
a random unstructured collection of amphiphillic molecules.

In the absence of any interactions that keep the membrane away from itself
\ref\HELII{W.~Helfrich, Z. Naturforsch. {\bf 33A} (1978) 305.},
there
is no lowest energy state since the defect strength $b$ need not be quantized
for
a single membrane.  Entropic repulsion will balance against the energy gain of
the defect, setting the value of $b$.  The entropic repulsion will scale
as $b^{-2}$ leading to some $b^*$ that will minimize the total free energy
above the lower critical chirality.  However, the entropic interaction
will scale as the area of the bilayer whereas the line energy will scale
like the length.  The detailed balance between entropic repulsion and
defect energy will thus be size dependent. It is also possible
that a preferred, spontaneous {\sl torsion} along the free edge
would lead to a specific value for $b$ \ref\HELIII{W.~Helfrich, J. Chem. Phys.
{\bf
85} (1986) 1085.}.  Thus it should be possible for chiral
lyotropic lamellae to exhibit isolated helicoidal fragments.

\newsec{Lamellar Melting and TGB Phases}

When membranes are stacked together, they can form multi-layered, lamellar
structures, similar to smectic-$A$ phases in thermotropic liquid crystals.
In highly swollen systems, the layer spacing $d$ is determined by entropic
repulsion \HELII .
As in thermotropic smectics, the  $L_\alpha$ phase excludes twist \dg .
Likewise,
a sufficiently strong chirality can favor the entry of twist into the
$L_\alpha$ phase
in the form of a TGB phase consisting of periodically repeated twist-grain
boundaries
composed of periodic arrays of screw dislocations \LR .  Though we know of no
report of
an experimental observation of this phase, we see no reason why it could not
exist.

Transitions in lyotropic systems are driven predominantly via changes in
concentration
rather than changes in temperature.  We can determine the surfactant
concentration, or
equivalently the layer spacing $d$, at which the $L_\alpha$ phase first becomes
unstable with respect to a proliferation of dislocations by calculating the
point
at which the total energy per unit length of dislocation first becomes
negative.
This calculation is analogous to the calculation of $H_{C1}$ in
a superconductor -- it establishes the mean-field instability of the $L_\alpha$
phase
to the formation of a TGB phase.  In the limit $d\gg\xi$, we find from \efrees\
the total
free energy per unit length of a dislocation is
\eqn\etotal{F={1\over d}\left\{
\gamma d^2\pi\ln\left(\lambda\over\xi\right) - K_2q_0d +\mu\vert
d\vert\right\},}
where the pre-factor of $1/d$ converts from free energy per turn to free
energy per unit length \ref\jt{We thank J.~Toner for discussions on this and
many other points.}.

Note that in \efreen , the elastic constants are {\sl membrane} elastic
constants,
related to the bulk elastic constants by a factor of $1/d$, the inverse layer
spacing, and
consequentially they do not change as the density changes (assuming that
density
changes are made most easily by changes in the layer spacing).
The density is inversely proportional to the layer spacing, and so, against
simple intuition, the energy of the defect (with $b=d$) decreases with
increasing
density.  This is a simple result: as the layer spacing increases, the elastic
energy of each distorted lamella grows since it must be distorted more to
connect the consecutive layers.  We thus predict that with {\sl increasing}
density the lamellar structure will be pocked with defects above a lower
critical chirality.  For a given value of $K_2q_0-\mu>0$ there will be a
critical
spacing $d^*$ (critical density $\rho^*\propto 1/d^*$)
below which (above which) the defects will penetrate.
{}From \etotal , we find:
\eqn\dstar{d^*={K_2q_0-\mu \over \pi\gamma\ln(\lambda/\xi)}.}
In general, a transition will occur when $d$ is reduced to {\sl twice} the
preferred Burger's vector
of the free membrane \elimits .
If fluctuations are unimportant, then $d^*$ is the layer spacing at which there
is
a second-order transition from the $L_\alpha$ to the TGB phase.

Of course, it is possible that fluctuations are strong enough to destroy any
TGB phase
that might form and that the $L_\alpha$ phase transforms to a chiral $N_L^*$
phase with twisted
orientational order like that of a cholesteric.  This phase is the liquid
crystal
analog of the melted vortex lattice of superconductors in a magnetic field
\ref\DRN{
D.R.~Nelson, Phys. Rev. Lett. {\bf 60} (1988) 1973;
D.R.~Nelson and H.S. Seung, Phys. Rev. B {\bf 39} (1989) 9153.}, which
may intervene {\sl between} the Meissner phase and the vortex phase of a
type-II
superconductor.
It is a cholesteric phase formed by melting the
TGB dislocation lattice while still retaining short-range smectic order.  Heat
capacity
measurements provide strong evidence of this phase in thermotropic systems
\NLS.
Possible phase diagrams based on the
analogy
between this system and high-$T_C$ superconductors are shown in Figure 2.

We propose that the layers melt via dislocation loop unbinding, as in the
smectic-$A$-to-nematic transition \NT .  Here, however,
the chiral bias will cause one handedness of screw dislocation to be preferred
over
the other.  This bias should change a second-order-like unbinding transition to
a first-order transition:  microscopic defect loops can no longer unbind
smoothly since a loop must contain an equal number of left and right handed
screws.  Either the defect loops will unbind and the dissident part of
the loop will move to the boundary, or, equivalently, the appropriate
dislocations
will nucleate from the boundary.

The chiral interaction also changes the simple unbinding
picture in another way:
from the above we can see that the free energy difference per unit length
between a wrongly-handed
screw and a correctly-handed screw will be:
\eqn\edf{\Delta F/L = 2\vert K_2k_0b\vert.}
We thus expect that, in the presence of thermal fluctuations, at the
unbinding transition of the defect loops, the largest size dislocation
loop will have a characteristic length
\eqn\eloop{L^*={\kbT\over 2 K_2\vert k_0\vert d}.}
As the chirality of the molecules becomes small, {\sl i.e.} as
$k_0\rightarrow 0$, the screw dislocations will grow to infinite size and will
proliferate, as in the usual non-chiral scenario.

We can estimate
the latent heat per molecule via a scaling argument.  If we
assume that without chirality there is a second-order smectic-$A$-to-nematic
transition (or nearly second order), then the transition will
be rounded out when the correlation length is on the order of $L^*$.
In this fluctuation dominated
``type II'' limit the transition from smectic order to isotropic order
would proceed via an intermediate cholesteric-like phase (either TGB or
$N_L^*$) as shown in Figure 2(a).  We believe that
the most
likely candidate for this phase is the $N_L^*$ phase \NLS\ of thermotropic
smectics \LubKam .
It is also possible that the isotropic phase intervenes before a TGB phase ever
forms, as
is shown in the phase diagram of Figure 2(b).
For fixed $d^*$ we may calculate $\lambda^2/\xi^2$.  In the limit that
$d^*\gg\xi^*$ and
$\lambda\gg\xi^*$, we
have:
\eqn\elox{{\lambda^2\over\left(\xi^*\right)^2}{1\over
\ln^3\left(\lambda/\xi\right)}
= {\pi^3K_2\gamma\over\left(K_2q_0-\mu\right)^3}.}
Thus we expect extreme type II behavior, {\sl i.e.} $\lambda\over\xi^*$ large
when $K_2q_0\approx\mu$ or when $K_2\gamma$ is very large.  The transition from
the lamellar to TGB phases would presumably be second-order while the
transition from the lamellar to the $N_L^*$ phase would be first-order due to
the
strong fluctuations.

In the latter case we can estimate the scaling
behavior of the free energy per unit volume at the transition by cutting off
a the transition at the size $L^*$.
Scaling
gives us for the entropy per unit volume:
\eqn\esca{\ell = T_C{\partial f\over\partial T} \sim
\left[(L^*)^{-1/\nu}\right]^{1-\alpha} \sim
\vert K_2k_0\vert^{(1-\alpha)/\nu},}
where $f$ is the free energy density, $t$ is the reduced temperature, $\nu$ is
the correlation length exponent and $\alpha$ is the usual specific heat
exponent of the nearby second-order transition.
Note that this result based on dislocation loop unbinding gives the {\sl same}
scaling
result in terms of $k_0$ as that
obtained previously via Landau theory \ref\TOM{T.C.~Lubensky, J. Phys. (Paris)
{\bf
C1}
(1975) 151.}.

Since the defects in the lamellar phase are necessarily accompanied by holes
in the layers, X-ray scattering should provide a view of the dislocation
unbinding transition leading to a chiral TGB or cholesteric phase.  As the
dislocation
loops grow to $L^*$, the scattering will show liquid like peaks at $q_\perp =
4\pi/(\sqrt{3}L^*)\propto k_0 d$, where $q_\perp$ is the wavevector
perpendicular to the layer
normals.  Though there will be a proliferation of holes in each
layer, only the holes coming from the {\sl same} dislocation loop will
be correlated, leading to an X-ray structure with peaks at the characteristic
inverse length scale.
This could be tested by changing the concentration of chiral
molecules
(or, equivalently, adjusting the ratio of left-handed to right-handed
enantiomers),
as well as by adjusting the layer spacing.

Finally we mention that it is possible for the lamellar stacks to be in
a ``type I'' limit.  In this case the melting from the stacked phase to
the isotropic phase will occur at one place: there will be no
intervening nematic or cholesteric-like phase.  We would expect this
transition to be first-order and would not occur via dislocation loop
unbinding.
Instead we would expect this transition to occur via dislocations nucleating at
the boundary.  We also note that in the zero chirality limit there are no known
transitions from lamellar structures to nematic structures.  In the chiral
case,
however, we have seen that chiral mesophases arise via the presence of chiral
topological defects.

\newsec{Acknowledgments}
It is a pleasure to acknowledge stimulating discussions with R.~Bruinsma,
S.T.~Milner,
D.~Nelson, P.~Nelson, T.~Powers,
D.~Roux, C.~Safinya and J.~Toner.
This work was supported in part by NSF Grants DMR94-23114.

\nfig\fone{A screw dislocation in a lyotropic lamellar phase.  Note that the
core
is devoid of all molecules and is filled with solvent.  In thermotropic
smectics,
the core is nematic and contains the mesogens.}

\nfig\ftwo{Possible phase diagrams for the lyotropic system as a function of
chirality $q_0$ and surfactant density $\rho$.  Solid lines indicate
first-order
transitions while dashed lines indicate second-order (or weakly first-order)
transitions.
(a) Lyotropic with four distinct phases: lamellar,
TGB, $N_L^*$ and isotropic.  (b) It is possible that the isotropic phase
intervenes {\sl before} the TGB phase occurs.  The {\sl phantom} TGB region is
shown
with hatched lines: it will never appear.}
\footatend\vfill\supereject\immediate\closeout\rfile\writestoppt
\baselineskip=14pt\centerline{{\bf References}}\bigskip{\frenchspacing%
\parindent=20pt\escapechar=` \input refs.tmp\vfill\eject}\nonfrenchspacing
\vfill\eject\immediate\closeout\ffile{\parindent40pt
\baselineskip14pt\centerline{{\bf Figure Captions}}\nobreak\medskip
\escapechar=` \input figs.tmp\vfill\eject}

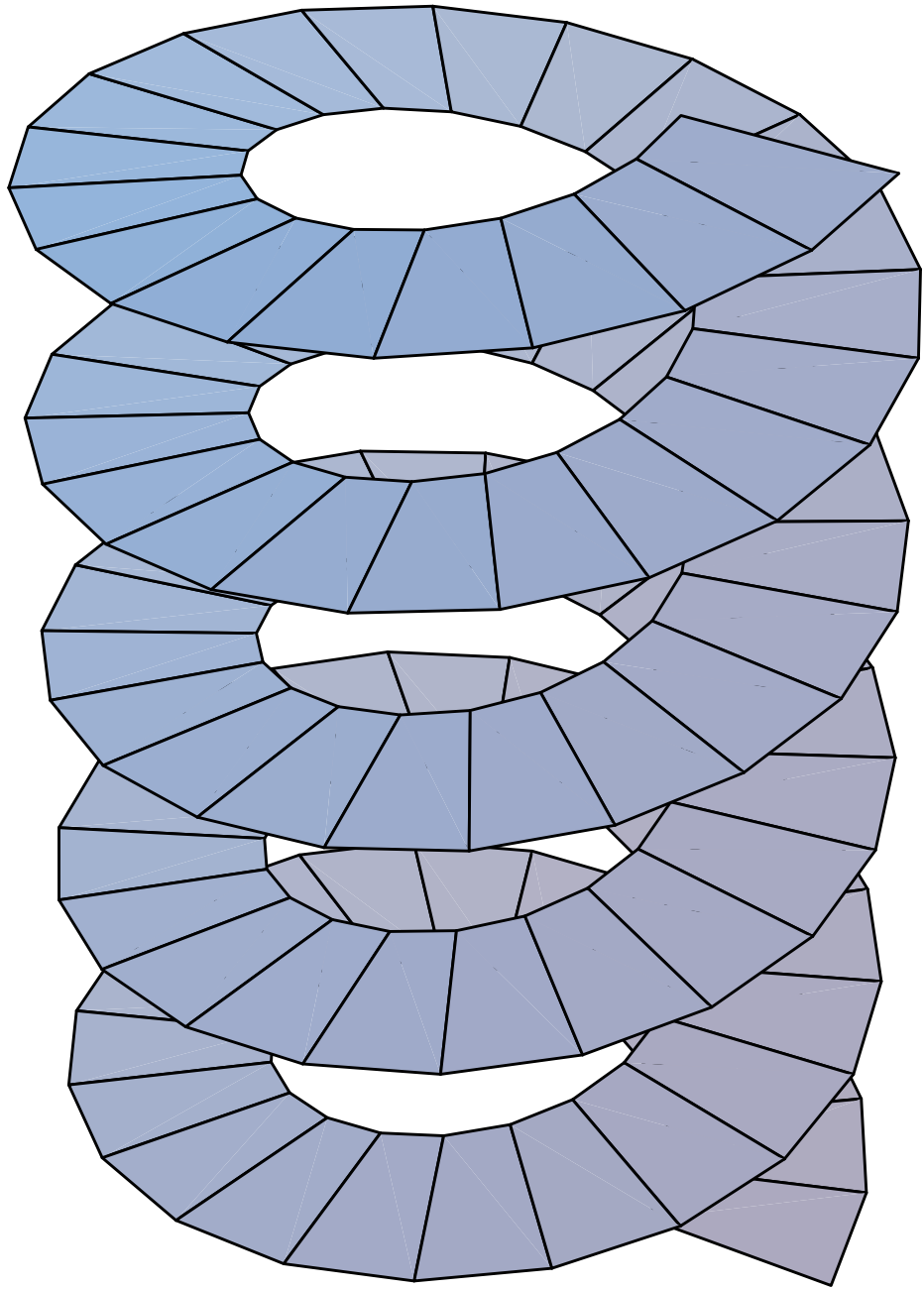
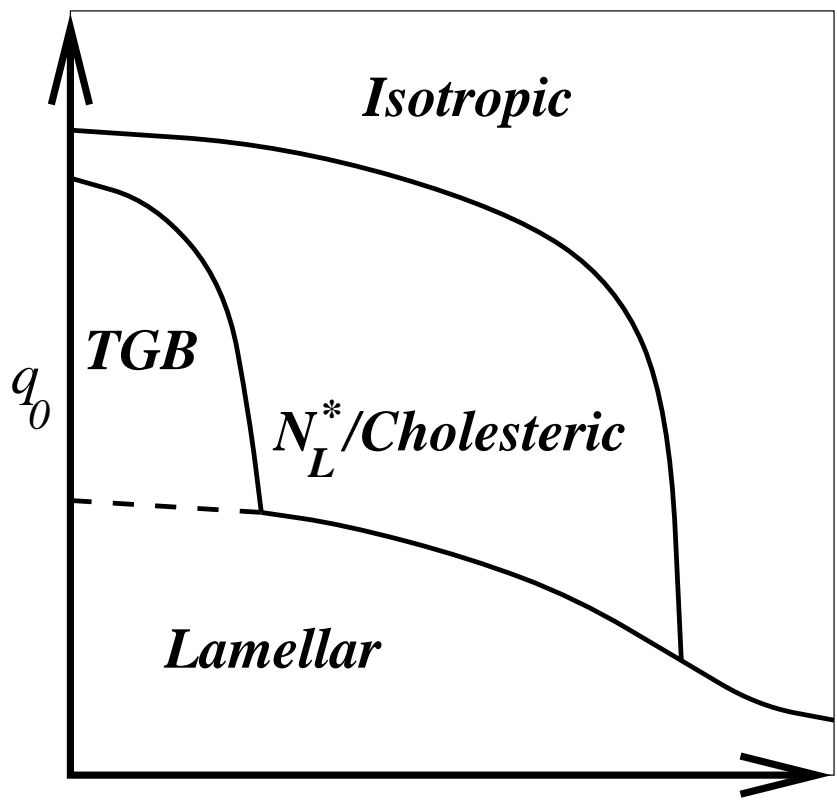
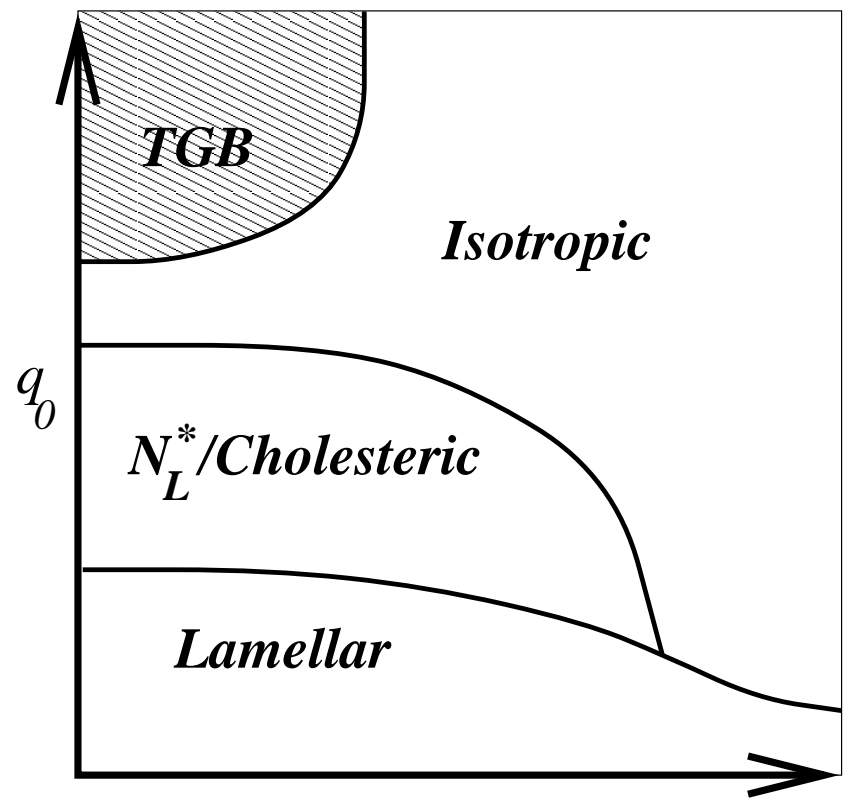
\bye